\documentclass{aastex63} 

\newcommand{\be}{\begin{equation}}
\newcommand{\ee}{\end{equation}}
\def\lta{\,\raise 0.3 ex\hbox{$ < $}\kern -0.75 em
 \lower 0.7 ex\hbox{$\sim$}\,}
\def\gta{\,\raise 0.3 ex\hbox{$ > $}\kern -0.75 em
 \lower 0.7 ex\hbox{$\sim$}\,}

\newcommand{\fhit}{f_{\rm surf}} 
\newcommand{\fseed}{f_{\rm seed}}
\newcommand{\mnrasalt}{{\it Mon. Notices Royal Astron. Society}}  
\newcommand{\apjalt}{{\it Astrophysical Journal}} 
\newcommand{\apjaltl}{{\it Astrophysical Journal Letters}} 
\newcommand{\aapalt}{{\it Astronomy \& Astrophysics}} 
\newcommand{\apssalt}{{\it Astrophysics and Space Science}} 

\begin{document}

\title{Transfer of Rocks between Planetary Systems: 
Panspermia Revisited}

\correspondingauthor{Fred C. Adams}
\email{fca@umich.edu}
\author[0000-0002-8167-1767]{Fred C.~Adams}
\affiliation{Department of Physics, University of Michigan, Ann Arbor,
  MI 48109, USA}
\affiliation{Department of Astronomy, University of Michigan, Ann
  Arbor, MI 48109, USA}
\author[0000-0003-4827-5049]{Kevin J.~Napier}
\affiliation{Department of Physics, University of Michigan, Ann Arbor,
  MI 48109, USA}


\begin{abstract}
Motivated by the recent discovery of interstellar objects passing
through the solar system, and by recent developments in dynamical
simulations, this paper reconsiders the likelihood for life bearing
rocks to be transferred from one planetary system to another. The
astronomical aspects of this lithopanspermia process can now be
estimated, including the cross sections for rock capture, the velocity
distributions of rocky ejecta, the survival times for captured
objects, and the dynamics of the solar system in both its birth
cluster and in the field. The remaining uncertainties are primarily
biological, i.e., the probability of life developing on a planet, the
time required for such an event, and the efficiency with which life
becomes seeded in a new environment. Using current estimates for the
input quantities, we find that the transfer rates are enhanced in the
birth cluster, but the resulting odds for success are too low for
panspermia to be a likely occurrence. In contrast, the expected
inventory of alien rocks in the solar system is predicted to be
substantial (where the vast majority of such bodies are not
biologically active and do not interact with Earth).
\end{abstract} 


\section{Introduction} 
\label{sec:intro}

The origin of life poses one of the most substantial --- and so far
largely inaccessible --- unresolved scientific questions. Although
this topic has many different facets, an important issue in
astrobiology is whether or not life can be transferred from one
planetary system to another. This type of transfer is generally known
as panspermia and has been studied extensively in previous work (e.g.,
\citealt{Melosh1988}; \citealt{Melosh2003}; \citealt{Napier2004};
\citealt{Wallis2004}; \citealt{adams2005}; \citealt{Valtonen2009};
\citealt{Wesson2010}; \citealt{Belbruno2012}; \citealt{Ginsburg2018};
\citealt{Lingam2018}). The recent discovery of the unbound
interstellar objects 'Oumuamua \citep{2017Natur.552..378M} and comet
Borisov \citep{2019ApJ...886L..29J} passing through our solar system
emphasizes that one part of the process, astronomical objects
traveling from one system to another, is not only readily realized,
but can also be observationally constrained. The other requirements
for the transfer of life, however, are expected to less common, and
most previous estimates suggest that the likelihood of transfer is
extremely low (see the above references). With a focus on the
astronomical aspects of the problem, this paper revisits the question
of panspermia. Although a great deal of previous work has been carried
out, the present treatment is motivated by recent observational input,
especially the detection of unbound interstellar objects, and by
recent numerical studies of the capture dynamics. We find that the
odds of successful transfer of life remain low, although significant
uncertainties remain.

\begin{figure}
\centering
\vskip1.0truecm
\includegraphics[width=0.65\textwidth]{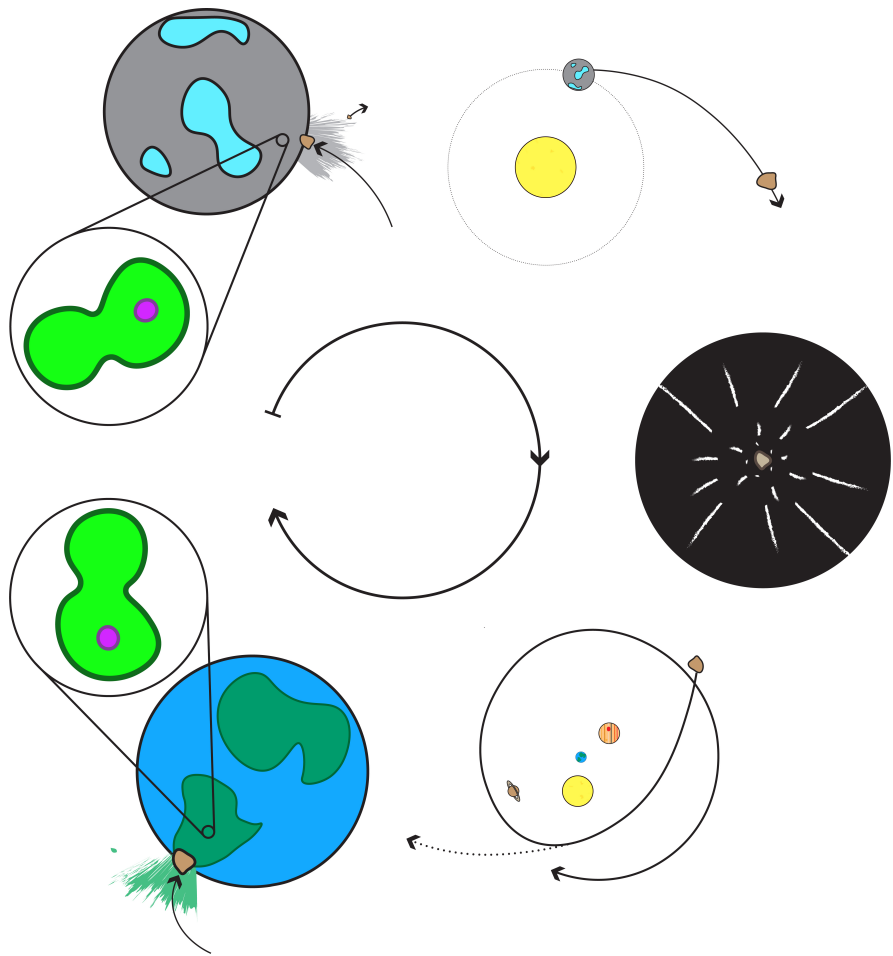}
\vskip1.0truecm
\caption{Schematic diagram showing the processes necessary for the  
successful transfer of life from one planetary system to another
(proceeding clockwise from the upper left to the lower left). Life
must first arise on a habitable planet. Rocks that contain living
biological material must then be removed from the planetary surface,
and then ejected from the planetary system. After traveling through
interstellar space, the rock must be captured by another planetary
system and delivered to the surface of a potentially habitable
planet. Finally, the biological material must thrive in its new
environment. }
\label{fig:schematic}
\end{figure}

The transfer of life between planetary systems requires a number of
astronomical and biological processes to take place. These steps are
illustrated in the schematic diagram of Figure \ref{fig:schematic}.
First, life must arise on a suitably habitable planet in order to
provide the seeds for other planetary systems. The habitable planet
must then be bombarded via impacts (or experience some other
cataclysmic event) in order to lift biologically active material off
the planetary surface and into interstellar space. These projectiles
then travel from one planetary system to another, and must be captured
by the recipient system(s). After capture into a bound orbit, the
biologically active body must eventually find its way onto the surface
of a habitable planet within its new system. Finally, the successful
transfer of life requires that life actually takes hold in its new
environment. The first and last of these processes are biological in
nature and their odds of occurrence are largely unknown. Of course,
Earth provides one example of life both arising and taking hold, but
an honest assessment of the odds remains out of reach. In contrast,
the dynamics of transfer can be calculated, and the corresponding odds
of success can be estimated. This present paper thus focuses on these
astronomical aspects of the problem.

As outlined in greater detail below, it is important to make the
distinction between different dynamical environments, namely within
clusters and in the field. Most stars form within embedded clusters
(\citealt{Lada2003,Porras2003}), and our Sun is thought to be no
exception
\citep{Hester2004,Zwart2009,Adams2010,Pfalzner2013,Parker2020}. Compared
to the field, such cluster environments are more conducive to the
transfer of rocky bodies between planetary systems
\citep{adams2005,Belbruno2012}. More specifically, the stellar density
within clusters is much higher than the field, by a factor of $\sim$
1000, and the relative speeds are lower, by a factor of $\sim$
40. Both of these properties enhance the capture of rocky
bodies. Working in the opposite direction, however, solar-type stars
typically spend more time in the field, by a factor of $\sim$ 100. As
outlined below, the advantages of the cluster environment outweigh the
loss of integration time. Many of the earlier estimates for the low
probability of rocky bodies seeding life in an alien solar system
(e.g., \citealt{Melosh2003}) correspond to current conditions in the
field (in the solar neighborhood). The consideration of the solar
birth cluster enhances the odds of transfer, but the odds remain low.

This paper is organized as follows. The first step is to consider the
ejection of rocky bodies from a planetary system that contains life
(Section \ref{sec:ejection}). Important considerations include the
total amount of material ejected, the distribution of ejection speeds,
and the size distribution of the rocky bodies. After rocks are
successfully ejected from a planet, they must be captured by another
planetary system, and the cross sections are considered next (Section
\ref{sec:capture}). Here we determine the cross section as a function
of asymptotic speed, with a focus on parameters appropriate for our
solar system (although the cross section has the same general form for
other systems). We then estimate the corresponding capture rates. Once
the rocks are captured, they must remain bound to the new planetary
system and eventually find their way onto the surface of a habitable
planet (Section \ref{sec:survival}). The results can be combined to
construct estimates for the optical depth for rock capture, the number
of successful panspermia events, and the standing population of alien
bodies in the solar system (Section \ref{sec:synthesis}). The paper
concludes (in Section \ref{sec:conclude}) with a summary of results
and a brief discussion of their implications.

\section{Ejection} 
\label{sec:ejection}

The first astronomical process of interest involves the removal of
rocks from the surface of a planet, and ultimately the ejection of the
rocks from the planetary system. On this point, we first note that
rocks are observed to be transported from one planet to another within
our solar system. For example, many meteorites found on Earth are
known to be of Martian origin \citep{McSween1985}, and the dynamics of
such transfer is well-studied (\citealt{Dones1999};
\citealt{Mileikowsky2000}; \citealt{gladman2005}).  Next we note that
the depth of the potential well from the surface of Earth (the escape
speed $v_e\sim11$ km/s) is comparable to, though somewhat smaller
than, the depth of the solar potential at $a=1$ au (where the orbital
speed $v_{\rm orb}\sim30$ km/s and escape speed $v_e\sim42$ km/s). We
thus expect that only a fraction of the rocky bodies that are removed
from planetary surfaces will be successful in leaving their planetary
systems, but such events should nonetheless take place.

\medskip
{\bf Mass in Ejected Rocks.} Regarding the total amount of mass that
can be ejected, simulations of the late stages of planet formation
(e.g, \citealt{Quintana2004}) and considerations of disk substructures
\citep{Rice2019} suggest that star/disk systems eject of order
$\sim1-10M_\earth$ of material during their early formative stages. 
We can parameterize this contribution by writing the total ejected 
mass in rocks in the form 
\begin{equation}
    M_R = f_R M_\oplus \,,
\end{equation}
where we expect the dimensionless parameter $f_R$ to be of order unity
(see also \citealt{Melosh2003,Dones1999,adams2005}).

Only a fraction of the mass $M_R$ will be ejected from the surface of
a particular planet in the system, and only a fraction of that mass
can possibly contain biologically active material. More specifically,
we are interested in rocks that contain some type of life-bearing
substance (spores, seeds, DNA) that can in principle start life on a
suitable new planet. The ejected mass of biologically infused rock 
can thus be written in the form 
\begin{equation}
    M_B = f_B M_R = f_B f_R M_\oplus\,,
\end{equation}
where we expect $f_B\ll1$. Although the value of $f_B$ is highly
uncertain, we can make an order of magnitude estimate as follows. 
Suppose that we have a habitable planet with properties
comparable to those of Earth. The mass of our biosphere is
approximately $10^{-10}M_\oplus$ (e.g., 
\citealt{biosphere}).\footnote{Note that this biomass landmass is 
dominated by land plants. In principle, seeds can be encased within 
rocks and transported through the panspermia mechanism, but 
experimental data is lacking. In addition, one should keep in mind 
that land plants have only existed for $\sim10\%$ of Earth history.} 
On one hand, we expect that only a fraction of the total mass of the
biosphere can be ejected; on the other hand, only a fraction of the
mass contained in the rocky ejecta is expected to contain biological
material. As a result, an approximate upper limit is given by
$f_B\lta10^{-10}$. As another estimate, previous authors find that
$\sim15$ biologically active rocks, with masses $\sim10^4$ g, should
be ejected from an Earth-like planet per year 
\citep{Melosh2003,Wallis2004}.  Over the age of the solar system, this
mass loss rate corresponds to a total ejected mass of
$\sim10^{-13}M_\oplus$ (so that $f_B\sim f_B f_R\sim10^{-13}$). The
ejection rate is likely to be higher at earlier stages of evolution,
so that this estimate is probably a lower limit (although life has
less time to develop at these earlier stages).  These considerations
thus indicate that the fraction of the ejected mass that is
potentially biologically activated lies in the range
\begin{equation}
    10^{-13} \lta f_B \lta 10^{-10} \,.
\end{equation}
This estimate has considerable uncertainty, and the range is
conservative. For example, if a planet develops life quickly and is 
subject to frequent impacts, then the ejected biomass could be a
larger fraction of the total. In the extreme limit, the integrated
biomass ejected over time could exceed the instantaneous mass of the
biosphere (this ordering depends, in part, on the carbon turnover time
-- see \citealt{carvalhais}).  Alternately, a planet could have a much
deeper biosphere than Earth, or more surface area covered by biomass,
which would increase the efficacy of the process.

\medskip
{\bf Distribution of Ejection Speeds.} In order for biologically
active rocks to be ejected from a planetary system, they must first
leave the planetary surface. Since the escape speed from an Earth-like
planet ($\sim11$ km/s) is somewhat smaller than the escape speed from
the orbital location within the system ($\sim42$ km/s), rocks released
from the planetary surface are likely to remain bound to the host
star. Over subsequent orbits, however, such bodies will interact with
other planets and can eventually be ejected from the planetary
system. If rocks are ejected by close encounters in the outer regions
of their solar systems, then the distribution of ejection speeds is
expected to have the approximate form
\begin{equation}
    F(v) = \frac{4v/v_p}{(1+v^2/v_p^2)^3}\,, 
    \label{eq:ejectdist} 
\end{equation}
where the velocity scale is determined by the depth of the stellar
potential well at the location of the planet \citep{Moorhead2005}. 
More specifically, the velocity $v_p \sim (GM_\sun/a_p)^{1/2}$, where
$a_p$ is the semi-major axis of the planetary orbit. Note that the
planet that ejects the rocks from the planetary system (and defines
$v_p$) is not necessarily the habitable planet of origin. In our solar
system, for example, Jupiter and Saturn are responsible for most
ejection events so that $v_p\sim10$ km/s, although other planetary
systems could have different architectures (see also
\citealt{Moorhead2005,adams2005}).

Note that equation (\ref{eq:ejectdist}) characterizes the distribution
of ejection speeds as the rocks leave their original planetary
systems. As they travel through interstellar space, the rocks will
eventually experience a series of distant encounters and will attain
the velocity distribution of the larger dynamical system. Here we are
interested in the birth clusters of planetary systems and the
field. In the cluster environment, the stars (and eventually the rocky
bodies) approach a Maxwellian velocity distribution with typical
dispersion $s\sim1$ km/s \citep{Lada2003}. Since much of the
distribution of equation (\ref{eq:ejectdist}) corresponds to higher
speeds, only the low-speed tail of the original distribution will be
retained within the cluster after the first crossing time. If we
define $u\equiv s/v_p$, then the fraction $f_s$ of the distribution
with speed $v<s$ is given by
\begin{equation}
    f_s = \frac{2u^2 + u^4}{1 + 2u^2 + u^4} \approx 0.075\,,
\end{equation}
where we have used $u=1/5$ to obtain the numerical value. The
resulting reduced population of rocks can eventually reach a
Maxwellian distribution, but will retain the distribution
(\ref{eq:ejectdist}) in the short term. For comparison, in the field
the stars have a Maxwellian distribution of speeds with typical
dispersion $s\sim40$ km/s \citep{Binney2008}. However, observations
indicate that the velocity dispersion increases with the age of the
stellar population. As a result, the ejected rocks in the field
eventually attain the same Maxwellian distribution as the star, but
the required time scale is expected to be greater than $\sim1$ Gyr
(e.g., \citealt{Haywood2013}).

\medskip
{\bf Distribution of Rock Sizes and Masses.} The above discussion
provides estimates for the total mass of ejected material from a
habitable planet. To determine the probability of biological transfer,
we need to specify the number of rocky bodies that could potentially
carry biological material from one planetary system to another.  The
rocky ejecta will be produced with a distribution of rock sizes and
masses.  Since this size distribution is expected to be a steeply
decreasing function of radius and hence mass (see below), the minimum
mass necessary for a rock to safely transport biological material
largely determines the total number of rocky bodies for a given total
mass.

The first step is to thus specify the minimum rock mass/size
necessary to transport biomass. A number of previous papers have
estimated the minimum rock mass and find that $m_{\rm min}\approx$
$10^4$ g (e.g., see \citealt{Horneck1993,Nicholson2000}). With this
minimum mass, the rock provides biological material with sufficient
shielding from interstellar cosmic rays and other hazards (see also
the discussion at the end of this section).

The mass distribution of the ejected rocks is expected to be nearly a
power-law, which can be written in the form 
\begin{equation} 
\frac{dN}{dm} = A m^{-p} \,,
\label{dndm} 
\end{equation} 
where $A$ is a normalization constant and the power-law index
$p=1-2$. More specifically, for the case where the size/mass
distribution is determined by collisional processes, one finds an
index $p\approx1.8$ \citep{dohnanyi1969,obrien1982,Napier2001}. 
For comparison, the observed distribution for objects striking the top
of the Earth atmosphere indicates an index $p\approx1.7$
\citep{Schroeder1991}. For indices in the expected range $1<p<2$, most
of the mass is contained in the (rare) largest objects, whereas most
of the bodies by number are the smallest objects. We thus introduce an
upper mass cutoff $m_2$ and a lower mass cutoff $m_1$. To a good
approximation, the normalization condition implies 
\begin{equation} 
A = \frac{(2-p) M_R}{m_2^{2-p}} \,,
\end{equation} 
and the total number of rocky objects per star is given by 
\begin{equation}
N_{R\ast} = \frac{(2-p)}{(p-1)} 
\left(\frac{m_2}{m_1}\right)^{p-1} \frac{M_R}{m_2} \,. 
\end{equation}  

If we take $M_R$ to be the total mass (in all the rocks ejected from a
given system), then $M_R\sim M_\oplus$ as discussed above. However,
the remaining parameters can vary for a wide range. For example,
\cite{Melosh2003} uses the values $m_1=10^4$ g, $m_2=0.1M_\oplus$ =
$6\times10^{26}$ g, and $p$ = 5/3, so that the total number of rocky
bodies per star becomes $N_{R\ast}\approx10^{16}$. For comparison,
\cite{Napier2001} uses the values $m_1=10^7$ g and $p$ = 11/6, so that
the number becomes $N_{R\ast}\approx 8 \times 10^{16}$. Since the
fiducial number of rocks is given by $N_0=M_R/m_1$, we define a
dimensionless parameter $f_N$ such that 
\begin{equation}
    f_N \equiv \frac{(2-p)}{(p-1)} \left(\frac{m_1}{m_2}\right)^{2-p}\,, 
\end{equation}
so that $N_{R\ast}=f_N N_0$. The number of biologically active rocks
produced per star is thus given by $N_{B\ast}=f_B f_N N_0$. Note that 
the number of rocks is sensitive to the value of the index $p$. Recent 
studies find a range of indices \citep{brown2002,suggs2014,ballouz2020}
with values somewhat larger than $p=1.7$. The steeper distributions 
imply larger numbers of small rocks (larger $f_N$). 

\medskip
{\bf Minimum Rock Size for Survival.} As outlined above, we need to
specify a lower cutoff to the mass distribution in order for the
integrated number of bodies to converge. In addition, in order to
survive transfer across interstellar space, biological material must
be encased within rocky bodies with a minimum mass. Here we take these
two mass scales to be coincident with a value of $m_1=10^4$ g. Notice
that by using the minimum mass for survival as the lower cutoff, we
are determining the number of rocky bodies large enough to carry
biological material.\footnote{We are thus neglecting the rocks that
  are too small to support life. Since most of the mass is contained
  in the largest objects, and most of the bodies by number reside in
  the smallest objects, this choice of lower mass cutoff does not
  affect the predicted number of biologically viable rocks for a given
  total mass.}

The value for the minimum mass $m_1=10^4$ g is advocated by
\cite{Melosh2003}, although a number of alternatives have been put
forth (see, e.g., \citealt{Napier2004}). Experiments show that UV
radiation can be lethal for spores and bacteria exposed to
interstellar conditions \citep{Horneck2001} and that a rocky casing of
a few centimeters is required for protection. The mass scale used here
($10^4$ g) corresponds to a larger size scale (compared to a few cm)
because of several additional considerations: First we note that the
rocky ejecta and their biological cargo must survive the impact
events that launch them from the surface of their original planet
\citep{Mastrapa2001,Horneck2008}, thus requiring somewhat larger
rocks. In addition, after being captured by another planetary system,
the rock must also survive the re-entry process onto the surface of
its new planet, where this latter process is likely to be the limiting
factor for survival \citep{Torre2010}.

\section{Capture Cross Sections} 
\label{sec:capture}

This section considers several mechanisms by which interstellar
objects, including those containing biological material, can be
delivered to our solar system (and other planetary systems). We begin
by reviewing the process of gravitational capture. We then consider
the case where interstellar objects collide directly with solar system
bodies and conclude by considering capture by gaseous circumstellar
disks.

\begin{figure}
    \centering
    \includegraphics[width=0.9\textwidth]{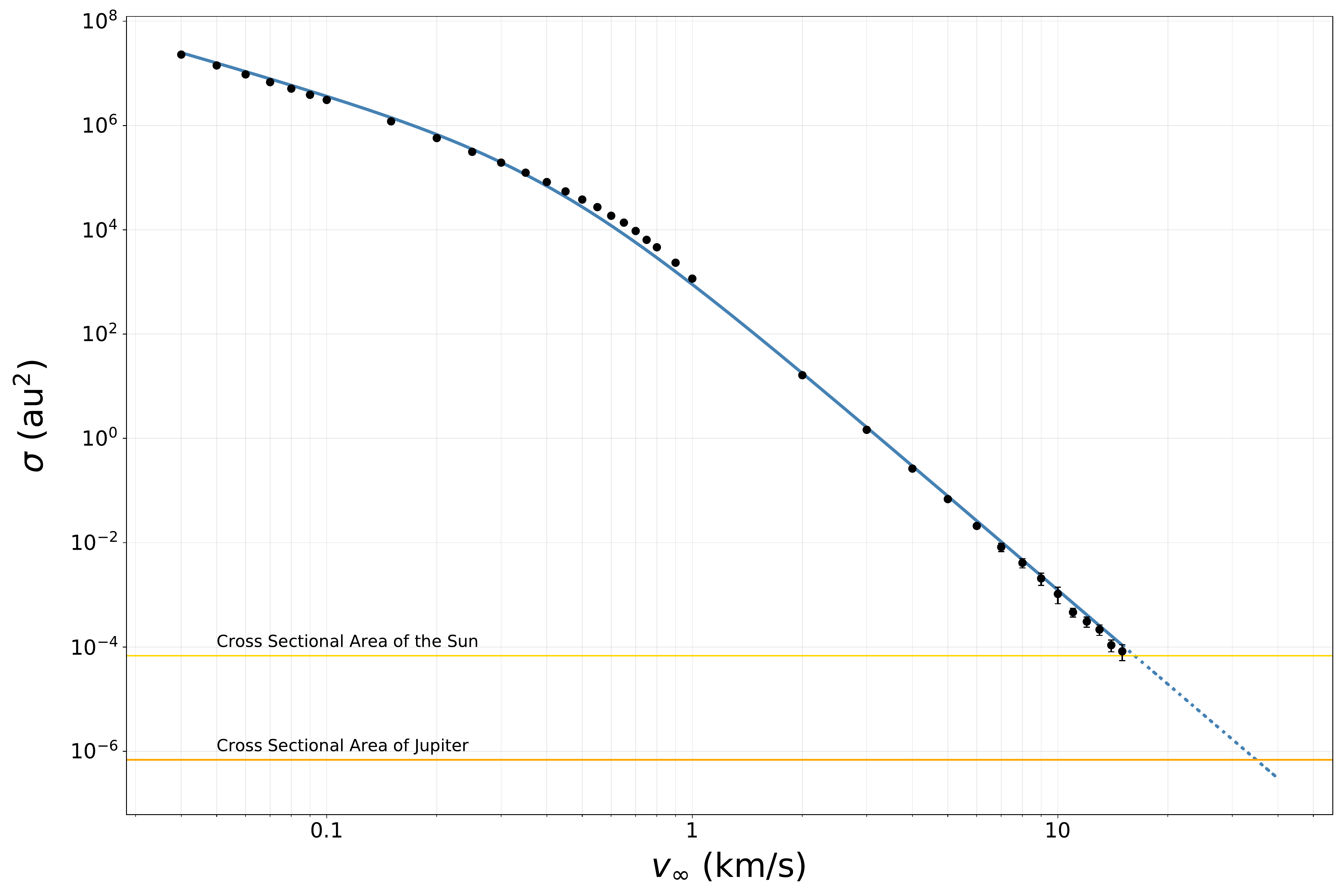}
    \caption{Capture cross section as a function of $v_\infty$. The
      black dots represent the data derived from the numerical
      simulations in \citealt{Napier2021}. The solid blue line is the
      best fit of equation (\ref{eq:cross-section}) to the numerical
      data. The dotted blue line is an extrapolation of the solid blue
      line from 15 km/s to 40 km/s. The horizontal yellow and orange
      lines represent the geometric cross sections of the Sun and
      Jupiter, respectively.}
    \label{fig:cross-section}
\end{figure}

\medskip
{\bf Capture by Dynamical Interactions.} The problem of gravitational
capture of interstellar objects is now relatively well-understood
(e.g., see the seminal paper by \citealt{Heggie1975}; see also the
more recent work of \citealt{Hands2020, Napier2021,Lehmann2021}). In
order for capture to take place, the incoming object must experience
the time-dependent gravitational potential induced by interactions
with the giant planets and the Sun. The capture dynamics can be
then broken into two regimes depending on the hyperbolic excess
velocities of the incoming objects. In the low-speed regime 
($v_\infty < 1$ km/s), the incoming bodies enter the sphere of
influence of the Sun. In the high speed regime ($v_\infty > 1$ km/s),
capture is dominated by close encounters with the giant planets. In
both regimes, the bodies exchange energy in a process that is
essentially the inverse of the well-known gravitational slingshot
effect. The orbital speed of the solar system body (either the Sun or
a giant planet) determines the maximum energy exchange, so that high
speed objects require interactions with the planets (which have higher
orbital speeds). Both analytic approximations
\citep{Heggie1975,Napier2021} and numerical simulations
\citep{Hands2020,Napier2021} show that the capture cross section
depends on the incoming asymptotic speed of the rock according to the
relation 
\begin{equation}
    \sigma(v_\infty) = \frac{\sigma_0}{(v_\infty/v_\sigma)^2 
\left[ 1 + (v_\infty/v_\sigma)^2\right]^2}\,,
    \label{eq:cross-section}
\end{equation}
where the best fit parameters for our solar system are given by
$\sigma_0 \approx 232,000$ au$^2$ and $v_\sigma \approx 0.42$
km/s. Equation (\ref{eq:cross-section}) enables the calculation of a
capture rate for any distribution of the incoming velocity $v_\infty$,
which we discuss in Section \ref{sec:synthesis}. Figure
\ref{fig:cross-section} shows the data for the capture cross section
derived from 500 million numerical experiments \citep{Napier2021}, as
well as the best fit from Equation (\ref{eq:cross-section}). The
numerical results span $\sim11$ orders of magnitude in capture cross
section and agree well with the analytic form over the entire range.
Note that the cross section is a sensitive function of $v_\infty$.
For other solar systems, the capture cross section has the general
form given by equation (\ref{eq:cross-section}), but with different 
values for the parameters $\sigma_0$ and $v_\sigma$. These values 
can be estimated using the analytic forms for the cross sections 
\citep{Heggie1975,Napier2021,Dehnen2021}. 

These data show that in the low-speed regime $\sigma(v_\infty) \propto
v_\infty^{-2}$, while in the high-speed regime, $\sigma(v_\infty)
\propto v_\infty^{-6}$. This velocity dependence implies that for
objects with high $v_\infty$, the capture cross section quickly
becomes exceedingly small. At $v_\infty = 15$ km/s, for example,
$\sigma(v_\infty)$ is comparable to the geometric cross section of the
Sun, and at $v_\infty = 40$ km/s, $\sigma(v_\infty)$ is comparable to
the geometric cross section of Jupiter. It is significant that there
is a maximum value of $v_\infty$ that an incoming body can have and
still be captured by our solar system where $v_{\rm max}\sim 40$
km/s. This speed limit is set by Jupiter's orbital speed (for a
derivation, see \citealt{Napier2021}). In principle, non-gravitational
effects such as atmospheric drag could allow faster bodies to be
captured, e.g., if they execute orbits that skim a planetary
atmosphere. However, such events are expected to be exceedingly rare
(even compared to direct collision events considered below), so that
the capture cross section for $v_\infty \gta 40$ km/s is effectively
zero.


\medskip
{\bf Capture by Direct Collision.} Another mechanism for capture
occurs through direct collisions with solar system bodies. As shown
above, the dynamical capture cross section for objects with asymptotic
speed $v_\infty \gtrsim 15$ km/s (35 km/s) is smaller than the cross
sectional area of the Sun (Jupiter). Furthermore, Jupiter (which is by
far the most effective planet at capturing interstellar rocks) can
only capture objects with $v_\infty \lesssim 40$ km/s. For cases where
the velocity distributions of the incoming bodies have large
dispersion, such in the field, direct collisions with a solar system
object --- rather than capture, followed by eventual collision --- can
be a more effective way of accumulating life-bearing rocks. The cross
section for directly landing on the surface of Earth is of particular
interest. The following discussion thus focuses on collisions with
Earth, but other solar system bodies (e.g., potentially habitable
moons of the giant planets) can be treated in analogous fashion.

In contrast to gravitational capture, direct collisions allow for
simple analytic approximations.  The capture cross section is the
geometric cross section of the body of interest with corrections for
gravitational focusing. For the case of Earth, the gravitational
focusing due to the Sun is a much larger effect than that due to Earth
itself. We first consider the cross section for an incoming rock to
reach a sphere of radius $r_0$ = 1 au centered on the Sun, i.e.,
\be
\sigma_1 = \pi r_0^2 \left( 1 + 
\frac{2GM_\odot}{r_0 v_\infty^2} \right)\,,
\ee
where $v_\infty$ is the asymptotic speed. Of course, the cross
sectional area of Earth is only a fraction $R_\oplus^2/4r_0^2$ of the
cross sectional area of the 1 au sphere. Note that we are assuming an
isotropic distribution of incoming bodies and a circular orbit. In
addition, the Earth will introduce its own gravitational focusing
factor, denoted here as $f_{g\oplus}$. With the inclusion of these two
effects, the cross section for direct capture by Earth can be written
in the form 
\be
\sigma = \frac{\pi}{4} R_\oplus^2 \left( 1 + 
\frac{2GM_\odot}{r_0 v_\infty^2} \right) f_{g\oplus}\,,
\ee 
where the final focusing factor $f_{g\oplus}$ is expected to be of
order unity. After the rock enters into the sphere of influence of
Earth, it will have a speed $v\sim v_{\rm orb}\sim30$ km/s, which
plays the role of the asymptotic speed for this second gravitational
focusing factor. As result, we find that $f_{g\oplus}\sim 1 +
(v_\oplus/v_{\rm orb})^2 \sim 1.13$, where the escape speed from Earth
$v_\oplus\sim11$ km/s (see the Appendix for a more detailed
calculation of Earth's gravitational focusing factor).

\medskip
{\bf Capture by Circumstellar Disks.} Another possible channel for the
capture of rocky bodies is provided by circumstellar disks
\citep{Napier2007,Brasser2007,Grishin2019,Napier2021}, which are
present during the earliest phases of evolution. The lifetime of
gaseous disks follows a nearly exponential distribution of the form 
\be
F(t) = \frac{1}{\tau} \exp\left[-t/\tau\right]\,,
\ee
where the time scale $\tau\sim5$ Myr \citep{jesus}. As a result, this
mechanism for capture operates only for early times -- when the solar
system is still within its birth cluster. If life is to be transferred
through this channel, then it must arise relatively early (on time
scales of a few Myr). Although timing issues are important for the
possible transfer of life, circumstellar disks can efficiently capture
rocks, as outlined below.

A rock passing through a gaseous disk will experience a drag force
that can be written in the form 
\be
F_D = A_{\rm eff} \rho_g v^2 \,,
\ee
where the effective area of the rock is related to the geometric cross
section according to $A_{\rm eff}=(C_D/2)\pi R^2$, where $R$ is the
rock radius and $C_D$ is the drag coefficient \citep{Landau1959}. In
order for a rock of mass $m_R$ to be captured, it must come to rest
within the nebula, where this criterion can be written in the from
\be
\frac{2A_{\rm eff}}{m_R} \Sigma > 
\log \left[ 1 + \frac{v_\infty^2 r}{2GM_\ast} \right] \,,
\ee
where $v_\infty$ is the asymptotic speed, and where the rock passes
through the disk at radius $r$ from the central star 
(\citealt{Napier2021}, Appendix B). Here $\Sigma$ is 
the surface density of the circumstellar disk (evaluated at the
location penetrated by the incoming rock). This expression assumes
that the rock passes vertically through the disk; corrections can be
applied to take into other angles of the trajectory and also the
relative velocity between the incoming rock and the gas due to orbital
motion in the disk. Note that within the birth cluster, the asymptotic
speed of incoming rocks $v_\infty\sim1$ km/s is generally much smaller
than the orbit speed at the locations of interest (where $v^2\sim
GM_\ast/r$), so that rocks will generally have incoming speeds
comparable to the local orbit speed when they encounter the disk. For
the sake of definiteness, we use the Minimum Mass Solar Nebula
\citep{Hayashi1981} as a benchmark model, where the surface density
has the form 
\be
\Sigma(r) = \Sigma_0 \left(\frac{r}{r_0}\right)^{-3/2}\,, 
\ee
where $\Sigma_0$ = 3000 g/cm$^2$ and $r_0$ = 1 au \citep{weidenschilling}.
With this profile, we can determine the capture condition and write
the result in terms of a cross section
\be
\sigma \approx \pi r_0^2 \left[ \frac{4 A_{\rm eff} \Sigma_0}{m_R} 
\frac{GM_\ast}{v_\infty^2 r_0} \right]^{4/5}\,.
\label{diskcross} 
\ee
Evaluating this expression for the minimum viable rock mass $m_R=m_1$
= $10^4$ g and for $v_\infty=1$ km/s (values appropriate for cluster
conditions), we find $\sigma\approx$ 69,000 au$^2$. This cross section
corresponds to a radial scale $\ell=(\sigma/\pi)^{1/2}\sim150$ au,
which is comparable to the expected outer radii of circumstellar
disks. As a result, rocks with the minimum mass of interest for
panspermia can be captured over essentially the entire disk area (but
only over a time span of 3 -- 5 Myr while the disk retains its gaseous
component).


\section{Survival Times and Transfer} 
\label{sec:survival}

In the previous section we considered the dynamical capture of
interstellar objects by our solar system. Although the capture and
survival of life bearing rocks must occur in order for life to be
transferred between planetary systems, most rocks do not find their
way onto the surface of a habitable planet. The vast majority of
captured bodies are ejected and those that survive the longest spend
most of their time in the outer regions of the solar system. As a
result, only a fraction $\fhit$ of the captured objects will reach the
surface of a habitable planet. This section reviews the dynamical
lifetime of captured interstellar objects and assesses the chances for
captured objects to reach the surface of Earth.

\begin{figure}
    \centering
    \includegraphics[width=0.9\textwidth]{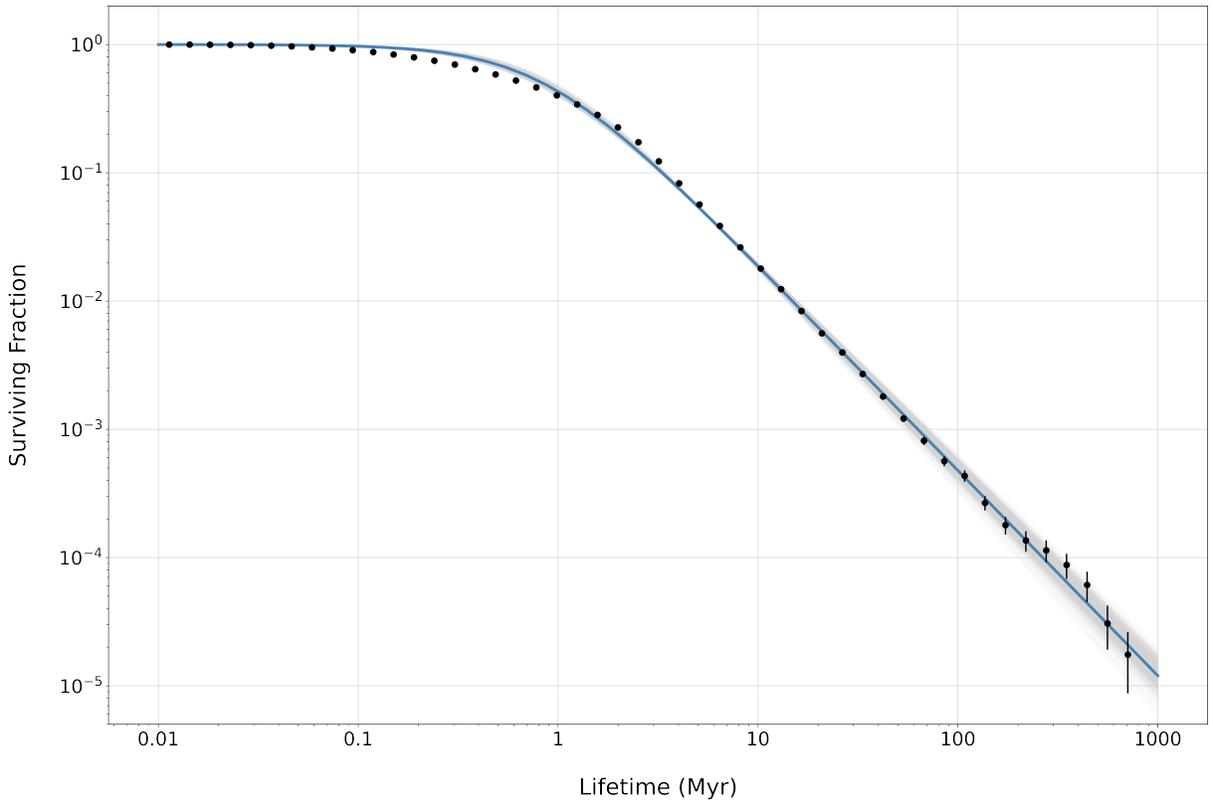}
    \caption{Surviving fraction of captured objects as a function of
      time. The black dots represent the data derived from the
      numerical simulations in \citealt{Napier2022}. The solid blue
      curve shows the best fit from equation (\ref{eq:survival}) to
      the numerical data.}
    \label{fig:survival}
\end{figure}

Most captured objects have lifetimes that are much shorter than the
age of the solar system. This behavior follows from the initial orbits
of the captured rocky bodies, which typically have semi-major axes
$a\sim1000$ au and periastra $q\sim10$ au \citep{Napier2021}. With
these orbital elements, captured objects tend to cross the orbits of
the giant planets (but not that of Earth). The continuing interactions
with giant planets, which have escape speeds larger than their orbital
speeds, can often lead to ejection. In a recent set of numerical
simulations \citep{Napier2022}, the orbits of newly captured objects
were integrated over the age of the solar system in order to determine
their lifetimes. The resulting fraction ${\cal F}(t)$ of remaining
bodies as a function of time since capture can be fit with a function
of the form
\begin{equation}
    {\cal F}(t) = \frac{1}{1 + (t/\tau)^{8/5}},
    \label{eq:survival} 
\end{equation}
where the time scale $\tau = 0.84$ Myr. Figure \ref{fig:survival}
shows the numerically determined survival function along with the
analytic approximation from equation (\ref{eq:survival}). The captured
objects survive for of order $\sim1$ Myr, with their population
steadily decreasing on longer time scales. Note that relatively few
objects surface on time scales $\sim1$ Gyr, so that the survival
fraction is not well-determined over longer times. Nonetheless, a
straightforward extrapolation of the numerical results suggest that
only a fraction $\sim10^{-6}$ (one in a million) of the captured
bodies can survive over the age of the solar system.

The numerical simulations show that during their residence time in the
solar system, captured bodies tend to have perihelia $q\gta10$ au. In
order to intercept an Earth-like planet, however, the orbit must have
$q\lta1$ au (where this condition is necessary but not
sufficient). Because of this mismatch in orbital elements, and because
of the small geometrical cross section, the fraction $\fhit$ of
captured bodies that impact the surface of a terrestrial planet is
expected to be extremely small. The numerical simulations used to
specify the survival function ${\cal F}(t)$ in equation
(\ref{eq:survival}) provide a working estimate of this quantity. The
simulations integrated the long-term behavior of $N_{\rm ob}$ =
276,691 captured objects, with most bodies being eventually ejected. 
About one third of the objects (specifically 107,812) have
orbits with perihelion $q<1$ au at some point while they remain
bound. These orbits have high eccentricities so that the objects
puncture the sphere of radius 1 au centered on the Sun as they make
their closest approach (and thus have a chance to collide with
Earth). Since the orbits are (nearly) isotropically distributed in
space, the fraction of orbits that correspond to planetary collisions
is proportional to $R_{\rm sph}^2/4r_0^2$, where $r_0$ = 1 au and
$R_{\rm sph}$ is the radius of the sphere of influence of Earth. The
radius $R_{\rm sph}$ $\approx r_0 (M_\oplus/M_\odot)^{2/5}\sim0.006$
au \citep{Bate1971}, so that less than $\sim1$ of the aforementioned
orbits can possibly intercept the planet. The number of collisions
could be higher because the bodies make multiple orbits. In the limit
where the orbits survive long enough to sample the full range of mean
anomaly for Earth, $\sim300$ of the objects would collide (this would
require the objects to live for $N\gg r_0/(2R_{\rm sph})\sim333$
orbits, which is more than typical). As a rough approximation, we can
use the survival half-life $\tau=0.84$ Myr to specify the typical
number of orbits $(\sim25)$ and find an optimistic estimate of
$\fhit\sim25/N_{\rm ob}\sim10^{-4}$.

In an independent estimate of the transfer fraction, \cite{Melosh2003}
performed a series of numerical simulations and found that a
comparable fraction $\fhit\sim10^{-4}$ of the captured objects reach
the surface of Earth over the age of the solar system. Rocky bodies
can also strike the surface of Jovian moons, which could represent
additional habitable environments (e.g., \citealt{Worth2013}). The
fraction of such impacts is about an order of magnitude smaller,
$\fhit\sim10^{-5}$. These results are thus comparable to those
described above. For the sake of definiteness, we take $\fhit=10^{-4}$
for the estimates of this paper (note that the results can easily be
scaled for different assumptions).

After arriving on the surface of a potentially habitable planet, any
given life-bearing rock will not necessarily be able to seed its new
world. Many difficulties are possible, e.g., the lack of a suitable
environment at the landing site, as well as the destructive effects of
passing through the atmosphere and crashing onto the planetary
surface. One expects many failed attempts. Unfortunately, the fraction
$\fseed$ of biologically active rocks that will successfully propagate
life on their new planet is largely unknown. Nonetheless, this
additional small parameter must be included in any assessment of the
likelihood of lithopanspermia taking place.

\medskip
\section{Synthesis}
\label{sec:synthesis} 

For a given background environment (the birth cluster or the field),
the solar system will accumulate rocks from interstellar capture at a
rate given by 
\be
\frac{dN}{dt} = n_R \langle \sigma v \rangle \,,
\ee
where $n_R$ is the number density of rocks and the angular brackets
denote an average of the cross section over the velocity
distribution. Note that the quantity $n_R$ corresponds to the total
number density of rocks, and that the number density of biologically
active rocks is smaller by a factor of $f_B\ll1$. The accumulated
number of rocks captured over a given time span is determined by the
integral of the capture rate, i.e.,
\be
N = \int \left[ n_R \langle \sigma v \rangle \right] dt\,, 
\ee
where the limits on the integral are implicit. In principle, all of
the quantities in the integrand can be time dependent. To a good
approximation, however, we can consider two distinct phases, i.e., the
first $\sim10-100$ Myr when the solar system resides in its birth
cluster and the subsequent $\sim4.5$ Gyr when the solar system resides
in the field. For each of these two phases, we can write the number of
captured rocks in the form
\be
N_x \approx N_{R\ast} \langle \sigma v \rangle_x
\left[ \int n_\ast dt\right]_x \equiv N_{R\ast} \tau_1 \,,
\ee
where the final equality defines $\tau_1$ as the optical depth for
capture per star. Here, the subscript $x$ denotes the phase (birth
cluster or field) and the integral is taken over the appropriate span
of time. The number of rocks per star $N_{R\ast}$ is discussed in
Section \ref{sec:ejection}. The other two factors are specified below.

\medskip
{\bf Velocity-averaged Cross Sections.} Using the cross sections
$\sigma(v)$ presented in Section \ref{sec:capture}, we can evaluate
$\langle\sigma{v}\rangle_x$ for the regimes of interest.  We start
with the cross section from equation (\ref{eq:cross-section}) for
dynamical capture. For the velocity distribution of equation
(\ref{eq:ejectdist}) appropriate for ejected objects (as expected 
in the birth cluster), we find the result 
\be
\langle\sigma{v}\rangle = \sigma_0 \frac{v_\sigma^3}{v_p^2} 
K(\eta) \,, 
\ee
where the dimensionless function $K(\eta)$ is defined such that 
$\eta\equiv v_\sigma/v_p$ and 
\be
K(\eta) \equiv \pi 
\frac{(1+\eta)(1+3\eta) + 3\eta^3/4}{(1+\eta)^4} \,.
\ee
Using parameter values appropriate for the birth cluster (and for
$v_p\approx10$ km/s), we find $\langle\sigma{v}\rangle_{bc}\approx$
540 au$^2$ km/s. Similarly, using a Maxwellian velocity distribution 
with dispersion $s$ (as expected in the field), we find 
\be
\langle\sigma{v}\rangle = \sigma_0 s
\left(\frac{v_\sigma}{s}\right)^2 
\sqrt{\frac{2}{\pi}} J(\mu) \,,
\ee
where the dimensionless function $J(\mu)$ is defined such that 
$\mu\equiv v_\sigma^2/(2s^2)$ and 
\be
J(\mu) \equiv \mu \left[ 1 - \mu {\rm e}^\mu E_1(\mu)\right] \,,
\ee
where $E_1$ is the exponential integral (see \citealt{Napier2021} for
a derivation).  Using parameter values appropriate for the field
($s=40\sqrt{2}$ km/s), we find $\langle\sigma{v}\rangle_{fd}\approx$
0.016 au$^2$ km/s.  

For the case of direct capture onto the Earth and the velocity
distribution of the birth cluster, the velocity averaged cross section
becomes 
\be
\langle\sigma{v}\rangle = \frac{\pi^2}{16} R_\oplus^2 v_p 
\left[ 1 + 6 \frac{GM_\odot}{r_0 v_p^2} \right] f_{g\oplus} \,.
\ee
Using the same values as before ($v_P=10$ km/s), we find
$\langle\sigma{v}\rangle_{dir-bc}\approx$ 
$7\times10^{-7}$ au$^2$ km/s. For
the velocity distribution of the field, with a Maxwellian form, the
corresponding expression takes the form 
\be
\langle\sigma{v}\rangle = \frac{\sqrt{2\pi}}{4} R_\oplus^2 s 
\left[ 2 + \frac{2GM_\odot}{r_0 s^2} \right] f_{g\oplus} \,.
\ee
The numerical value (for $s$ = $40\sqrt{2}$ km/s) becomes
$\langle\sigma{v}\rangle_{dir-fd}\approx$ 
$2\times10^{-7}$ au$^2$ km/s. Note
that the (averaged) direct capture cross sections have roughly
comparable values (within a factor of $\sim3$) in both the birth
cluster environment and in the field.

For the case of disk capture, we only need to consider the parameters for the birth cluster since the disk lifetime is shorter than the residence time in the cluster. We can write the disk-capture cross section from equation (\ref{diskcross}) in the form 
\be
\sigma(v) = \sigma_0 \left( \frac{v}{v_1} \right)^{-8/5} \,,
\ee
where $\sigma_0\approx69,000$ au$^2$ and where we have defined $v_1=1$ km/s. The velocity averaged cross section thus takes the form 
\be
\langle\sigma{v}\rangle = 4 \sigma_0 v_p \left( \frac{v_1}{v_p} \right)^{8/5} \int_0^\infty \frac{u^{2/5} du}{(1+u^2)^3} = \sigma_0 v_p \left( \frac{v_1}{v_p} \right)^{8/5} \frac{39\pi}{25(1+\sqrt{5})}\,, 
\ee
which can be evaluated to find $\langle\sigma{v}\rangle\approx$ 26,000 au$^2$ km/s. The disk is thus much more efficient in capturing passing rocks than either dynamical or direct capture. On the other hand, the other two channels can take place over the entire age of the solar system, whereas the disk lifetime is only a few Myr ($\sim1000$ times shorter). 

{\bf Integrated Number Densities.} The next quantity of interest is the stellar number density integrated over residence time. The stellar number density in the solar neighborhood is measured to be $n_\ast\approx0.1$ pc$^{-3}$ \citep{Binney2008}, and can be considered as constant over the age of the solar system (which corresponds to the integration time). For the field, we thus find $\int n_\ast dt \approx 450$ pc$^{-3}$ Myr.

For the solar birth cluster, a number of studies have placed
constraints on this integral, which is bounded from above by the
requirement that the solar system is not overly disrupted during its
residence time. Various considerations include the undisturbed nature
of the planetary orbits, the observed edge of the Kuiper Belt, and
the narrow spread of mutual inclination angles of solar system bodies
(e.g., see \citealt{Adams2001,Zwart2009,Li2015,Pfalzner2013,
Parker2020,Batygin2020,Moore2020}; and the review of
\citealt{Adams2010}). Although the lack of disruption implies an upper
bound on $\int n_\ast dt$, many solar system properties can be
explained if the Sun lived within its birth cluster for an extended
span of time (typically, 10 -- 100 Myr). For example, the observed
elevated abundances of short-lived radioactive nuclei (such as
$^{26}$Al) could be provided by a supernova explosion within the birth
cluster (e.g., see \citealt{Cameron1995}). The requirement that the
cluster provides such a supernova implies a lower bound on the
integral $\int n_\ast dt$ \citep{Adams2001}. Although many
uncertainties remain, a self-consistent picture of the early solar
system --- while it remains in its birth cluster --- indicates that
$\int n_\ast dt \sim 10^4$ pc$^{-3}$ Myr.

{\bf Capture Optical Depth.} Using the above considerations, we can
evaluate the optical depth for rock capture. The optical depth for a
given rock to be captured by the solar system is determined by the
quantity $\tau_1$ = $\langle\sigma{v}\rangle$ $\left[\int n_\ast
  dt\right]$.  These values, along with the aforementioned components,
are listed in Table \ref{tab:values}. The table also lists the total
number of rocks $N_{\rm cap}=N_{R\ast}\tau_1$ that are predicted to be
captured during the two phases of interest. To evaluate this quantity,
we use $N_{R\ast}=10^{16}$ rocks per star and assume that only 10\% of
the ejected rocks remain in the birth cluster (see Section
\ref{sec:ejection}). Next, we provide an estimate for the number
$N_{\rm bio}$ = $f_B N_{\rm cap}$ of biologically active rocks that
could be captured by the solar system. The values listed in the table
assume that the fraction $f_B=3\times10^{-12}$, which corresponds to
the geometric mean of the allowed range. The total number $N_{\rm
  pan}$ of expected panspermia events is determined by the fraction of
the captured rocks that cross Earth's orbit and land on its surface,
and by the fraction $\fseed$ of those events that lead to the
successful propagation of life. As a result, $N_{\rm pan}$ =
$\fseed\fhit N_{\rm bio}$. For the entries in Table \ref{tab:values},
we use the fraction $\fhit=10^{-4}$ (see Section \ref{sec:survival}
and \citealt{Melosh2003}), but leave the fraction $\fseed$
unspecified. Although we expect $\fseed\ll1$, we have no means of
estimating its value.

\begin{table}[h!]
\centering
\bigskip
\bigskip 
\centerline{\bf Rock Transfer Estimates}
\bigskip
\begin{tabular}{cccccc}
quantity & Birth Cluster & Field & Direct & Disk & units \\
\hline
$\langle\sigma{v}\rangle$ & 540 & 0.016 & $3\times10^{-7}$ & 26,000 & au$^2$ km/s \\ 
$\int n_\ast dt$ & $10^4$ & 450 & 1500 & 25 & pc$^{-3}$ Myr \\ 
$\tau_1$ & $10^{-4}$ & $3\times10^{-10}$ & $10^{-14}$ & $10^{-5}$ & -- \\ 
$N_{\rm cap}$ & $10^{11}$ & $2\times10^{6}$ & 100 & $10^{10}$ & -- \\ 
$N_{\rm bio}$ & 0.3 & $5\times10^{-6}$ & $3\times10^{-10}$ & 0 &  -- \\ 
$N_{\rm pan}$ & $3\times10^{-5}\fseed$ & $5\times10^{-10}\fseed$ & $3\times10^{-14}\fseed$ & 0 & -- \\ 
\end{tabular}
\medskip 
\caption{Cross sections, optical depths, and expected numbers of
events for rock capture in the solar birth cluster and in the
field. The optical depth $\tau_1$ determines the probability of a
given rock being captured by the solar system (note that $\tau_1$ 
is given by the product of the entries in the first two lines, when 
converted into the proper units). Here, $N_{\rm cap}$ is the expected
number of captured rocks and $N_{\rm bio}$ is the expected number of
biologically active rocks captured. Only a fraction of those rocks
cross Earth's orbit and have a chance to seed life. To estimate the
expected number $N_{\rm pan}$ of panspermia events (final row), we
have taken this fraction to be $\fhit\sim10^{-4}$. The number of
panspermia events must also be multiplied by the unknown (but probably
small) fraction $\fseed$, which specifies the likelihood for life to
thrive in its new environment. }
\label{tab:values}  
\end{table}

\bigskip 

In Table \ref{tab:values} we also include estimates for the transfer
of rocks onto the surface of Earth by direct capture. In this
scenario, where interstellar rocks directly strike the Earth's
surface, the velocity averaged cross section $\langle\sigma{v}\rangle$
is almost the same for both the birth cluster and the field. In this
context, the gravitational focusing factor (in the cluster
environment) nearly compensates for the reduced speed. For this
estimate we thus combine the integrated number density $[\int n_\ast
  dt]$, including the reduction factor in the birth cluster because
most of the ejected rocks are unbound. The net result is that the
number of rocks captured by directly striking the surface of Earth is
about $10^4$ times smaller than the total number of rocks captured by
the entire solar system from the field. However, all of the rocks that
strike the Earth in direct capture events reach the surface of a
habitable planet, by definition, so the reduction factor $\fhit$ does
not apply. As result, approximately the same number of panspermia
events result from direct capture and from dynamical capture in the
field (whereas the number of events from dynamical capture in the
birth cluster is far greater).

For completeness, Table \ref{tab:values} also includes the expected
capture events by the circumstellar disk. Since the disk lifetime is
only $3-5$ Myr, these interactions take place in the birth
cluster. This short time span results in the reduced effective value
of the integral $\int n_\ast dt$. In addition to the shorter
integration time, the number density of rocks is smaller because
ejection takes places over the entire cluster lifetime. Although a
large number of rocks are captured by the disk ($\sim10^{10}$, roughly
comparable to the number from dynamical capture in the birth cluster),
no biologically active rocks are expected due to the short lifetimes.

Putting all of the factors together, we can write the total number of
expected panspermia events in the form 
\be
N_{\rm pan} = \sum_{x} \left[ N_{R\ast} \fseed \fhit 
f_B \langle\sigma{v}\rangle \int n_\ast dt \right]_x\,,
\label{sum} 
\ee
where the sum is taken over the different channels for capture. Note
that in principle all of the efficiency factors could vary with the
capture mechanism. For the estimates given in Table \ref{tab:values},
the sum of equation (\ref{sum}) is dominated by capture events in the
birth cluster, and even this quantity $(\sim10^{-5}\fseed)$ is much
smaller than unity. This number could be even smaller if life cannot
arise during the time that the solar system spends in the cluster. In
that case, the number of events drops to $\sim10^{-9}\fseed$, with
comparable contributions from direct capture by Earth and from
gravitational capture by the solar system followed by subsequent
transfer to the planetary surface. Finally we note that the galaxy
contains $\sim10^{11}$ stars and that the fraction of stars harboring
Earth-like planets $\eta_\oplus\approx0.1$ \citep{Petigura2013}. As a
result, the number of panspermia events for the entire galaxy is
estimated to be $\sim10\fseed$.

It is important to note that the input parameters required to evaluate
equation (\ref{sum}) are uncertain. The number of rocks per star
depends sensitively on the index $p$ of the power-law distribution of
rock sizes (see equation [\ref{dndm}]), so that $N_{R\ast}$ can vary
by several orders of magnitude. In addition, the value of the factor 
$f_{\rm seed}$, which specifies the likelihood for life to survive and
flourish, is rather poorly constrained. We also note that additional
channels of transfer can be considered. For example, if life arises
rapidly, then rocky bodies can be captured while the surface area of
solid bodies was dominated by planetesimals instead of planets, so
that the cross section for direct capture (via impact) would be
larger. Moving away from the particular properties of our solar
system, the transfer of terrestrial life to other planetary systems
would be more likely if those systems had architectures with less of
an impedance mismatch between capturing planets like Jupiter and
Saturn and accretors like the Earth
\citep{tremaine1993,bonsor2012,wyatt2017}. In addition, a sizable
fraction of stars reside in binary systems, which have larger capture
cross sections than our solar system (compare the results of
\citealt{adams2005} with those of \citealt{Napier2021}). We also note
that planetary systems can live in different galactic environments
with larger stellar density (like the galactic bulge) and that the
optical depth for capture is directly proportional to $n_\ast$.

The results found here indicate that the best chance of transferring
life-bearing rocks occurs with the birth cluster. The main uncertainty
for this scenario is the time required for life to arise on any planet
within the cluster, since the cluster lifetime in only $\sim100$ Myr.
On one hand, one argument put forth in favor of panspermia is that life
began too quickly on Earth to be the result of internal processes like
`random chemical interactions' \citep{Wesson2010}.  On the other hand,
if life can arise rapidly on other planets in the birth cluster, the
underlying necessity of panspermia is diminished. In any case, the
time when life began on Earth remains highly uncertain, varying from
just after the Moon-forming impact at $\sim4.5$ Ga to $\sim3.8$ Ga 
\citep{sleep2018}.

\section{Conclusion}
\label{sec:conclude} 

This paper presents an updated assessment for the probability of rock
transfer between planetary systems, including the question of whether
life can propagate through this mechanism. We find that the
probability of transferring life from one planetary system to another
is quite low. On the other hand, the chances of transferring rocks
between systems is quite high. A more specific summary of our results
is given below, followed by a discussion of uncertainties and possible
future directions.

\medskip
{\bf Summary of results.} Successful panspermia events are predicted
to be exceedingly rare. The number of potentially life-bearing rocks
that are captured per solar system is found to be of order unity in
the birth cluster and nearly a million times smaller in the field. In
order for life to be successfully transferred from one planetary
system to another, however, the biologically active rocks must also
make their way onto the surface of the new habitable planet (e.g.,
Earth) and life must subsequently thrive. The first process is known
be a low probability event, whereas the odds of the second are
essentially unknown. In addition, in order for the enhanced estimates
for the birth cluster to be applicable, life must arise on a
relatively short time scale ($t\ll100$ Myr --- while the sun resides
within the cluster). If life takes $\sim1$ Gyr to arise (a common
assumption in astrobiology; \citealt{Lunine2005,Scharf2009}), then the
expected number of panspermia events is 
$N_{\rm pan}\sim10^{-10}\fseed$ where $\fseed\ll1$. 

Although the solar system has resided in the field for much longer
than in its birth cluster, the cluster environment dominates the
process of rock transfer. The lower relative speeds allow for much
larger capture cross sections and the higher densities act to further
increase the capture rates. These effects more than compensate for the
reduced residence time in the birth cluster ($t\lta100$ Myr) compared
to the field ($t\sim4.5$ Gyr).

Captured rocks have a well-defined distribution of (initial) orbital
elements which continue to evolve afterwards, and generally lead to
ejection from the system. The fraction of surviving bodies follows a
well-defined function, which has the form $f(t)\sim t^{-8/5}$ at late
times $t\gg1$ Myr (see Figure \ref{fig:survival}, equation
[\ref{eq:survival}], and \citealt{Napier2022}). This steeply declining
survival function reduces the chances of captured objects striking the
surface of Earth.

Combining the capture rates with survival rates, one can estimate the
total mass in alien rocks that currently resides on bound orbits
within the solar system. These results (see \citealt{Napier2022})
indicate that the mass captured from the birth cluster and that remains
today is given by $M_{RC}\sim10^{-9}$ $M_\oplus$; for comparison, the
standing population of rocks captured from the field has estimated
mass $M_{RF}\sim7\times10^{-14}M_\oplus$. For the size distribution of
rocks used here (Section \ref{sec:ejection}), the corresponding
numbers of captured rocks are approximately $N_{RC}\sim10^7$ from the
birth cluster and $N_{RF}\sim70$ from the field.

\medskip 
{\bf Discussion.}  The estimates of this paper are subject to a number
of uncertainties. Significantly, the astrophysical parts of the
problem are now relatively well understood. We have good working
estimates for the capture cross sections, the survival times for
captured objects, and the distributions of velocities for ejected
rocks. These quantities have been determined by both numerical
simulations and by supporting analytic calculations. Additional
astrophysical elements of the problem have been estimated, including
the dynamical properties of the birth cluster, the collision rate of
captured bodies on planetary surfaces, and the distribution of rock
sizes.  The most important remaining uncertainties are those derived
from biology. In particular, we would like to know the time required
for life to arise, how often habitable planets actually produce life
forms, and the likelihood of life gaining a foothold after crash
landing into a new environment. These latter quantities combine to
determine the fraction of rocks $f_B$ that contain biological material
and the fraction $\fseed$ of rocks that can seed a new planet after
being delivered to its surface. The factors $(f_B,\fseed)$ thus
encapsulate the largest uncertainties regarding panspermia estimates.

One important bottleneck is that life must arise spontaneously at
least once --- otherwise the panspermia mechanism cannot operate.
Current estimates suggest that the spontaneous development of life is
rare \citep{Lunine2005,Scharf2009} --- otherwise panspermia would
not be necessary.  Since the transfer of life between planetary
systems is also quite rare (see Table \ref{tab:values}), the key
question is which of these processes is more common (see
\citealt{Davies2003} for an extended discussion of this issue).

Most discussions of panspermia focus on estimates for the probability
that Earth was seeded by biological material from afar. Given that
Earth is the one place in the galaxy where life is known to exist,
however, it also makes sense to consider whether or not life from our
solar system has propagated elsewhere. The results of this paper (see
also references herein) can be used to make such estimates. However,
the cross sections and survival times depend on the architecture of
the planetary system capturing the rocks, and the results presented
here are carried out for our solar system. Future work could thus
explore how varying planetary configurations will affect the
results. As one line of inquiry, binary star systems are expected to
have enhanced capture cross sections \citep{Li2015}, but also have
somewhat reduced real estate for habitable planets \citep{David2003}.

Although the focus of this work is to consider astrophysical
processes, specifically the ejection and dynamical capture of rocks
through gravitational interactions, some authors have considered the
directed propagation of life (see, e.g.,
\citealt{Crick1973,Oneill1974,Jones1976,Walters1980,Djosovic2018,Lingam2018};
and many others). The directed propagation of life across the galaxy
is a far more efficient process than the dynamical processes
considered here (see also \citealt{Vukoti2012,Djovs2019}).

Although clusters provide a means of enhancing transfer of rocks
between constituent solar systems, the dense environment also presents
additional hazards. In particular, cluster environments have intense
ultraviolet radiation fields and elevated levels of cosmic radiation.
For example, the distribution of possible UV fluxes is wide
\citep{Fatuzzo2008}, but the median value is typically $\sim3000$
times larger than the interstellar value (which is
$\sim1.6\times10^{-3}$ erg cm$^{-2}$ s$^{-1}$). As result, the cluster
environment can be more destructive to life forms than the field. On
the other hand, the cluster environment also provides another
opportunity for the transfer of life: If the cluster itself, or one of
its constituent planetary systems, can capture life-bearing rocks from
outside, then life can readily be transferred to other members of the
cluster. In this context, the cluster acts as an amplification
mechanism for the transfer of life.

The results of this paper show that while the transfer of rocks
between planetary systems is common, the transfer of life is predicted
to be rare. In addition to the scenario outlined here, however, other
possible channels exist \citep{Burchell2004}. For example, even if
captured bodies do not cross Earth's orbit and impact its surface,
out-gassing can disseminate microbes into interplanetary space
\citep{Hoyle1999}. This process could take place early in solar system
history, when many of the rocks are captured and while the planets are
still forming. In such a chaotic environment, biological material can
be spread throughout the field of rocky debris \citep{Narlikar2003}.
Provided that these seeds of life live long enough, they could be
swept up by Earth, Mars, or perhaps even Europa as they trace through
their orbits. Although standard lithopanspermia is highly inefficient,
one should keep in mind that alternate scenarios could play a role in
the propagation of life through the galaxy.

\appendix

\section{Gravitational Focusing by the Earth}


In Section \ref{sec:capture} we used an estimate to compute Earth's
gravitational focusing factor, as the correction from using the full
treatment is rather small. In this appendix we provide a full analytic
treatment for the gravitational focusing factor as a function of the
incoming body's orbital parameters (see also
\citealt{opik1951,shoemaker1982}).

Earth's gravitational focusing gives it an effective radius of
\begin{equation}
    R_{\rm eff} = R_\earth 
\left( 1 + \frac{v_{\rm esc}^2}{v_{\infty\earth}^2}\right)^{1/2},
\end{equation}
where $v_{\infty\earth}$ represents the hyperbolic excess velocity of
the incoming object as it enters into a hyperbolic orbit about
Earth. In general, calculation of $v_{\infty\earth}$ depends on the
incoming body's semi-major axis ($a$), eccentricity ($e$), and
inclination ($i$). All of these quantities are all well-defined for
both hyperbolic and elliptical orbits, so the following formalism
should be generally applicable. We begin by considering the incoming
body's specific energy and angular momentum, defined by
\begin{equation}
    E = -\frac{GM_\sun}{2a} \qquad {\rm and} \qquad 
J^2 = GM_\sun a (1 - e^2).
\end{equation}
It follows that the radial velocity of the incoming body with respect
to the Sun is given by
\begin{equation}
    v_{r\sun}^2 = 2E - \frac{J^2}{r^2} + \frac{2 G M_\sun}{r} 
= \frac{2 G M_\sun}{r} - \frac{G M_\sun a}{r^2}(1 - e^2) - \frac{GM_\sun}{a}.
\end{equation}
The non-radial velocity component of the incoming orbit is determined
by conservation of angular momentum so that
\begin{equation}
    v_{\Omega\sun}^2 = \frac{J^2}{r^2} = \frac{G M_\sun a}{r^2}(1 - e^2)
\end{equation}
We can then break this quantity into azimuthal and longitudinal
components as follows:
\begin{equation}
    v_{\varphi\sun}^2 = v_{\Omega\sun}^2 \cos^2{i} 
\qquad {\rm and} \qquad 
v_{\vartheta\sun}^2 = v_{\Omega\sun}^2 \sin^2{i}.
\end{equation}
We now make the assumption that Earth is moving on a circular orbit
with $i = 0$. In this approximation, the incoming rock's radial and
longitudinal velocities are equivalent in the Earth and Sun frames. 
The velocity components can then be written in the Earth frame as
\begin{equation}
    v_{r\earth}^2 = v_{r\sun}^2 \qquad v_{\varphi\earth}^2 
= (v_{\varphi\sun} - v_\earth)^2 \qquad v_{\vartheta\earth}^2 
= v_{\vartheta\sun}^2.
\end{equation}
Then the relative velocity of the encounter is given by
\begin{equation}
    v_{\infty\earth}^2 = v_{r\earth}^2 + 
v_{\varphi\earth}^2 + v_{\vartheta\earth}^2.
\end{equation}
If we work through some algebra and make the substitutions $r = a_\earth$ and $v_\earth^2 = GM_\sun/a_\earth$, this expression can be written as
\begin{equation}
    v_{\infty\earth}^2 = v_\earth^2 \left[3 - \frac{1 - e^2}{\xi^2} - 2\xi\cos{i}\right],
    \label{vinfearth} 
\end{equation}
where we have defined $\xi \equiv \sqrt{(a/a_\earth)(1-e^2)}$. We see
that the velocity scale is set by $v_\earth \approx 30$ km/s. This
expression is minimized when the body is moving in the same direction
as Earth, and maximized when moving toward Earth head-on. Including
the gravitational focusing factor, Earth's effective cross section
becomes enhanced by the factor 
\begin{equation}
f_{g\oplus} = \left( 1 + \frac{v_{\rm esc}^2}{v_{\infty\earth}^2}\right)\,,
\end{equation}
where $v_{\infty\earth}$ is given by equation (\ref{vinfearth}).

\acknowledgments

We would like to thank Konstantin Batygin, Juliette Becker, and David
Gerdes for useful discussions. We also thank the two referees, Luke
Dones and an anonymous reviewer, for many constructive comments. This
material is based upon work supported by the National Aeronautics and
Space Administration under Grant No. NNX17AF21G issued through the SSO
Planetary Astronomy Program and by the National Science Foundation
under Grant No. AST-2009096.

\end{document}